\documentclass[conference]{IEEEtran}

\usepackage[utf8]{inputenc}
\usepackage[T1]{fontenc}
\usepackage[english]{babel}

\usepackage{cite}
\usepackage[pdftex]{graphicx}
\usepackage[cmex10]{amsmath}
\usepackage{amssymb}
\usepackage{algpseudocode}
\usepackage{color}
\usepackage{array}
\usepackage[font=footnotesize,skip=5pt]{caption}
\usepackage{url}
\usepackage[font=footnotesize]{subcaption}

\newtheorem{definition}{Definition}

\newcommand{\privapi}{\textit{Promesse}} 
\newcommand{\wfm}{$\mathcal{W}4\mathcal{M}$} 

\begin{document}

\author{\IEEEauthorblockN{Vincent Primault\IEEEauthorrefmark{1},
Sonia Ben Mokhtar\IEEEauthorrefmark{1},
Cédric Lauradoux\IEEEauthorrefmark{2} and 
Lionel Brunie\IEEEauthorrefmark{1}}
\IEEEauthorblockA{\IEEEauthorrefmark{1}Université de Lyon, CNRS\\
INSA-Lyon, LIRIS, UMR5205, F-69621, France\\
\{vincent.primault,sonia.ben-mokhtar,lionel.brunie\}@liris.cnrs.fr}
\IEEEauthorblockA{\IEEEauthorrefmark{2}INRIA\\
cedric.lauradoux@inria.fr}}

\title{Time Distortion Anonymization for the Publication of Mobility Data with High Utility}

\maketitle

\begin{abstract}
An increasing amount of mobility data is being collected every day by different means, such as mobile applications or crowd-sensing campaigns.
This data is sometimes published after the application of simple anonymization techniques (e.g., putting an identifier instead of the users' names), which might lead to severe threats to the privacy of the participating users. 
Literature contains more sophisticated anonymization techniques, often based on adding noise to the spatial data.
However, these techniques either compromise the privacy if the added noise is too little or the utility of the data if the added noise is too strong.
We investigate in this paper an alternative solution, which builds on time distortion instead of spatial distortion.
Specifically, our contribution lies in (1)~the introduction of the concept of time distortion to anonymize mobility datasets (2)~\privapi{}, a protection mechanism implementing this concept (3)~a practical study of \privapi{} compared to two representative spatial distortion mechanisms, namely Wait For Me, which enforces $k$-anonymity, and Geo-Indistinguishability, which enforces differential privacy.
We evaluate our mechanism practically using three real-life datasets.
Our results show that time distortion reduces the number of points of interest that can be retrieved by an adversary to under 3~\%, while the introduced spatial error is almost null and the distortion introduced on the results of range queries is kept under 13~\% on average.
\end{abstract}


\section{Introduction}

The widespread adoption of location-aware devices such as smartphones, GPS navigation systems or GPS-enabled watches has dramatically increased the quantity of mobility data that is being continuously collected.
This data is either sent \textit{online} by the users when querying location-based services (e.g., finding nearby restaurants, friends) or is collected by applications and then published/sold \textit{offline} to third parties (e.g., GPS traces sold for marketing purposes) or as part of crowd-sensing campaigns (e.g., publication of jogging/cycling traces, publication of pictures with location metadata on Flickr).

However, collecting and sharing mobility data raises serious privacy concerns.
Among the known threats is the extraction of users'\textit{points of interest}~\cite{Gambs11} (or POIs), which can be defined as places where individuals regularly spend some time, e.g., home, work, a cinema or a mall.
By studying the semantics of these places, it is possible to infer sensitive knowledge like religious or political preferences, for example if an individual often goes to a worship place or a political party's headquarter.
Learning users' POIs can ultimately lead to learn about their real identity with a good accuracy~\cite{Krumm07}.
Mobility data can also be used as an input of next place prediction algorithms, enabling to guess where users are likely to be in the next hour~\cite{Gambs12} as well as in four years~\cite{Sadilek12}, which opens the door to harmful threats such as house robbing.

Nevertheless, mobility data is still very valuable.
Publishing such information allows researchers to perform real-time traffic predictions, to find out interesting mobility patterns or to discover social and economic tendencies.
A number of research efforts have thus been carried out in the last few years to protect users' location privacy.
While effective mechanisms exist for the protection of privacy when a user opportunistically sends her current location to a location-based service (e.g., by using two non-colluding servers~\cite{Guha12}), it is still an open issue to publish mobility traces of a set of users in a privacy-preserving manner.
The major challenge in the latter case is due to the regularity of users' mobility habits.
Indeed, as users go to the same places at similar times of the day/days of the week, simple clustering techniques applied on published traces allow to infer users' POIs.

To address this issue, a classical solution is to alter user locations (e.g.,~\cite{Andres13,Abul08}).
However, according to the amount of added noise, this may also alter the utility of the published data.
Indeed, the introduced spatial distortion may prevent the exploitation of protected traces for some well-known use cases such as transportation mode detection (e.g.,~\cite{Zheng08b}).
An alternative solution that one may think of is the adoption of time distortion instead of spatial distortion.
In this paper, we investigate this alternative and introduce our main contribution, \privapi{}, which is, at the best of our knowledge, the first mechanism that aims at hiding users' POIs by distorting time.
Specifically, \privapi{} hides users' POIs by: (1)~smoothing the users' speed along their trajectories and (2)~blurring the start and end points of these trajectories to make them less easily identifiable.

A preliminary version of this work has been presented in~\cite{Mapomme15a}.
We present in this paper a detailed version along with a thorough evaluation.
We practically study the effectiveness of \privapi{} compared to two representative mechanisms relying on spatial distortion, namely Wait For Me~\cite{Abul10}, which guarantees $k$-anonymity, and Geo-Indistinguishability~\cite{Andres13}, which guarantees differential privacy.
Our evaluation, performed using three real-life datasets, shows that the number of retrieved POIs with \privapi{} is under 3~\%, which is comparable to what the other mechanisms can achieve.
In the same time, \privapi{} provides no spatial error, while the other mechanisms' error ranges from 24 to 70,000 meters.
 
The remaining of this paper is structured as follows.
We first present related works in Section~\ref{sec:related} before introducing the problem statement in Section~\ref{sec:statement}.
We then present our mechanism in Section~\ref{sec:detailed}.
We further present our experimental evaluation in Section~\ref{sec:evaluation}.
Finally, we conclude the paper and present our future research directions in Section~\ref{sec:conclusion}.

\section{Related work}
\label{sec:related}

There is a rich literature about privacy-preserving data publication of mobility traces.
We categorize existing work depending on the privacy guarantee they offer.

\textbf{\boldmath$k$-anonymity.}
In 2002 Sweeney introduced the concept of $k$-anonymity.
The main idea behind this concept is that, for each quasi-identifier of a published dataset, there must be at least $k$ persons with the same quasi-identifier.
This allows to hide each person among $k-1$ other persons.
$k$-anonymity has been extended to the field of location privacy with spatial cloaking~\cite{Gruteser03}.
In this context people report to be in an approximate area instead of reporting their exact location.
By doing this they make their location indistinguishable among all locations contained in their area (if an adversary has no background knowledge) and they are anonymous among all other persons in the same area.

When people move, they essentially move from one place to another, which is often a POI.
The list of these places can been considered as a quasi-identifier, which can be protected with a $k$-anonymity guarantee.
Terrovitis et al.~\cite{Terrovitis08} presented a model where they know the background knowledge of an adversary and use it to suppress a set of locations from trajectories, taking into account the impact both in terms of privacy and utility.
Yarovoy et al.\cite{Yarovoy09} tackled the problem of creating optimal anonymization groups for moving objects, which unlike traditional databases may not be disjoint.
However, these approaches tackle a slightly different problem than ours.
They are interested in anonymizing a sequence of POIs and not in anonymizing whole trajectories of users, which is what we want to do.
Other mechanisms consider trajectories as a whole instead of individual locations or places.
Abul et al. proposed \textit{Never Walk Alone}~\cite{Abul08}, whose idea is to guarantee that at every instant there is at least $k$ users walking at a given distance of the others, thus creating cylinders within which users move.
This mechanism has been later improved by \textit{Wait for Me}~\cite{Abul10} (or \wfm{}).
The latter removes some constraints about the input dataset and scales to large datasets.
Their mechanism was tested against two datasets, a synthetic one and a real-life one.
The approach was shown to perform well with the synthetic dataset but having more difficulties to maintain a correct utility with the real-life dataset.

\textbf{Differential privacy.}
Differential privacy is a more recent concept introduced by Dwork~\cite{Dwork06} defining a formal and proven privacy guarantee.
The idea is that an aggregate result over a database should be almost the same whether or not a single element is present inside the database.
In other words, the addition or removal of one single element shall not change significantly the probability of any outcome of aggregate functions.
One manner to provide differential privacy is by adding calibrated Laplacian noise to each component of a query result.

Differential privacy has been used by Jiang et al.~\cite{Jiang13} to protect ships' trajectories.
Endpoints of trajectories are preserved while intermediate locations are altered.
This results in a large distortion of trajectories and consequently in a difficult trade-off between privacy and utility.
Andres et al.~\cite{Andres13} introduced the concept of \textit{Geo-Indistinguishability} (or Geo-I) which is a generalization of differential privacy specifically suited for location privacy.
They propose to achieve Geo-I by adding a calibrated noise drawn from a two-dimensional Laplace distribution.
Two use cases are studied in their paper, one of them being the the anonymization of statistical geolocated datasets.
We have previously studied this mechanism in~\cite{Mapomme14} and shown that it is not suitable to anonymize entire mobility datasets, because POIs can still be retrieved.
Chen et al.~\cite{Chen12} anonymized public transportation usage data, which can be seen for each user as a sequence of places (metro/bus stations) she went to.
They built a method to anonymize such data in a differentially private way and evaluated their mechanism by studying the impact of anonymization on range queries and sequential pattern mining.
Like other $k$-anonymity works, their mechanism is focused on the anonymization of sequences of records.
Acs et al.~\cite{Acs14} proposed a mechanism to anonymize spatio-temporal densities datasets, which reports counts of active users within small areas for given time windows.
Such data can be obtained for instance from call data records that are gathered by mobile phone operators.
The authors proposed an approach that adapts to the original data in order to guarantee differential privacy with the highest possible utility.
Counting users is one interesting thing to do with mobility data, but we want to publish entire trajectories and allow more mining tasks to be performed.

\section{Problem statement}
\label{sec:statement}
\noindent We present in this section the privacy and utility objectives we target in this paper.

\subsection{Utility concerns}
We aim at the \textit{publication} of \textit{fine-grained} and \textit{spatially accurate} mobility datasets.
By fine-grained datasets we mean a set of records, each one being a location (i.e., a point on the Earth) associated with a (virtual) user identifier and a timestamp.
Indeed, publishing fine-grained datasets allows analysts to freely implement their data mining tasks with the tools and languages they want and run them on the published data.
Moreover, there are use cases where such datasets are required. For example, researchers working on delay-tolerant networks test their algorithms with real-life datasets (e.g.,~\cite{Hui08}).
Another example is the case of transportation mode detection.
In this context, state-of-the-art algorithms need to extract a number of information from mobility traces such as speed, acceleration~\cite{Zheng08b} or proximity to rail lines/bus stops~\cite{Stenneth11}, which is not possible using alternative solutions such as interactive querying, where data analysts are restricted to a pre-defined query language provided by the data owner (e.g., ~\cite{Pelekis11}), or the publication of pre-aggregated datasets (e.g., ~\cite{Acs14}).

\subsection{Privacy challenges}
In terms of privacy, we focus on \textit{hiding the users' POIs} that could be inferred by an attacker.
As defined in~\cite{Gambs11}, POIs are important places that a user regularly visits and in which she spends a given amount of time.
POIs have been shown to be very sensitive because they allow to infer knowledge such as where one lives, her work occupation, her hobbies, her political or religious preferences, etc. 
urther, it has been shown in~\cite{Golle09} that even the simple pair home/work can lead to a re-identification of a large number of users.
We hence focus in this paper on the protection of users' POIs that we assess using a state-of-the-art POI extraction attack presented in~\cite{Hariharan04}.

\subsection{The need of a new mechanism}
State-of-the-art protection mechanisms suitable to publish fine-grained mobility datasets in a privacy-preserving way focus on spatial distortion, i.e., they alter the spatial component of each record to protect the privacy of users.
We present in the next section an alternative solution based on temporal distortion and analyse the pros and cons of each approach in the rest of the paper.

\begin{figure*}
        \centering
        \begin{subfigure}[b]{0.3\textwidth}
                \includegraphics[width=5.2cm]{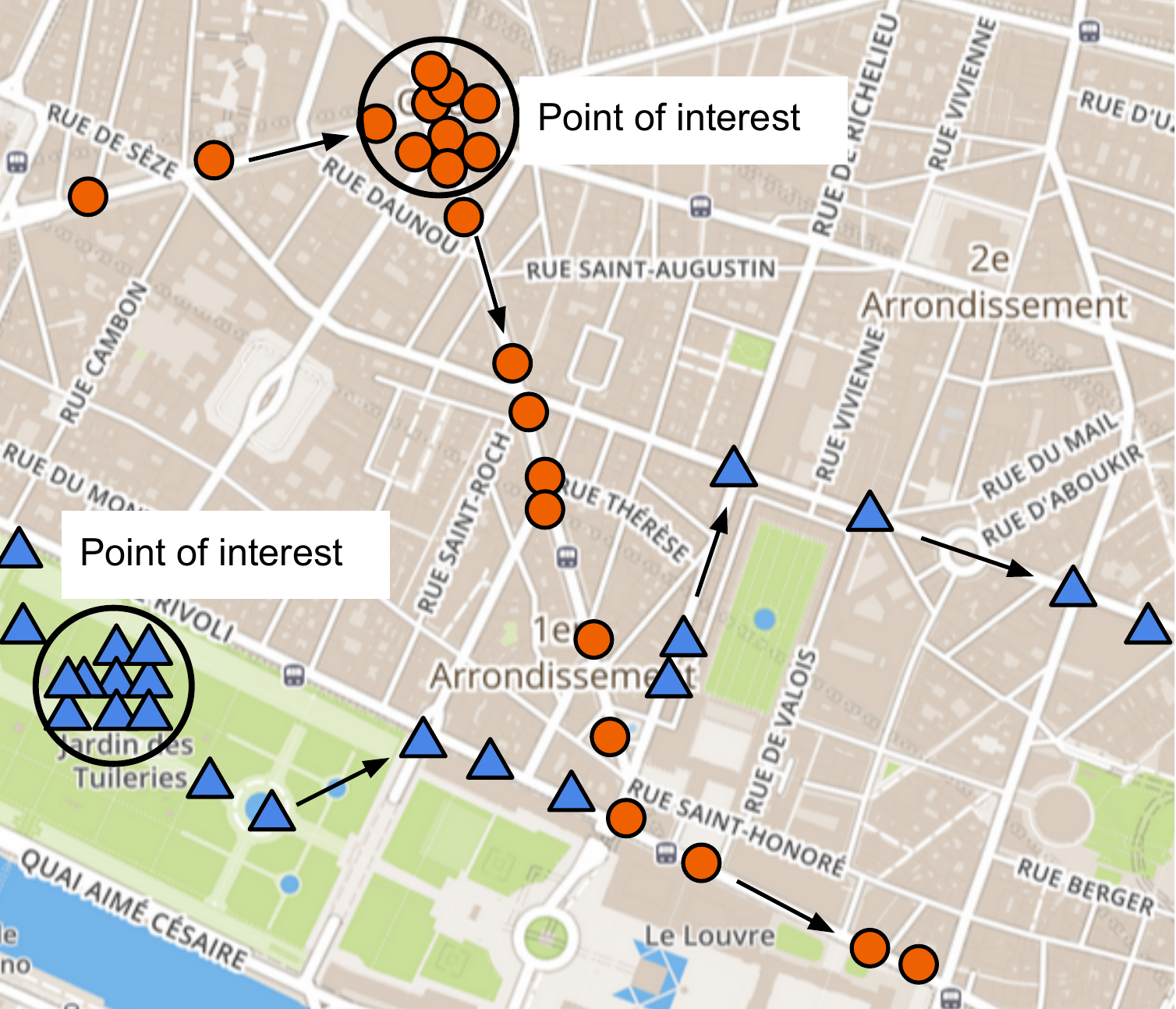}
                \caption{Original dataset.}
                \label{fig:original}
        \end{subfigure}%
        ~ 
        \begin{subfigure}[b]{0.3\textwidth}
                \includegraphics[width=5.2cm]{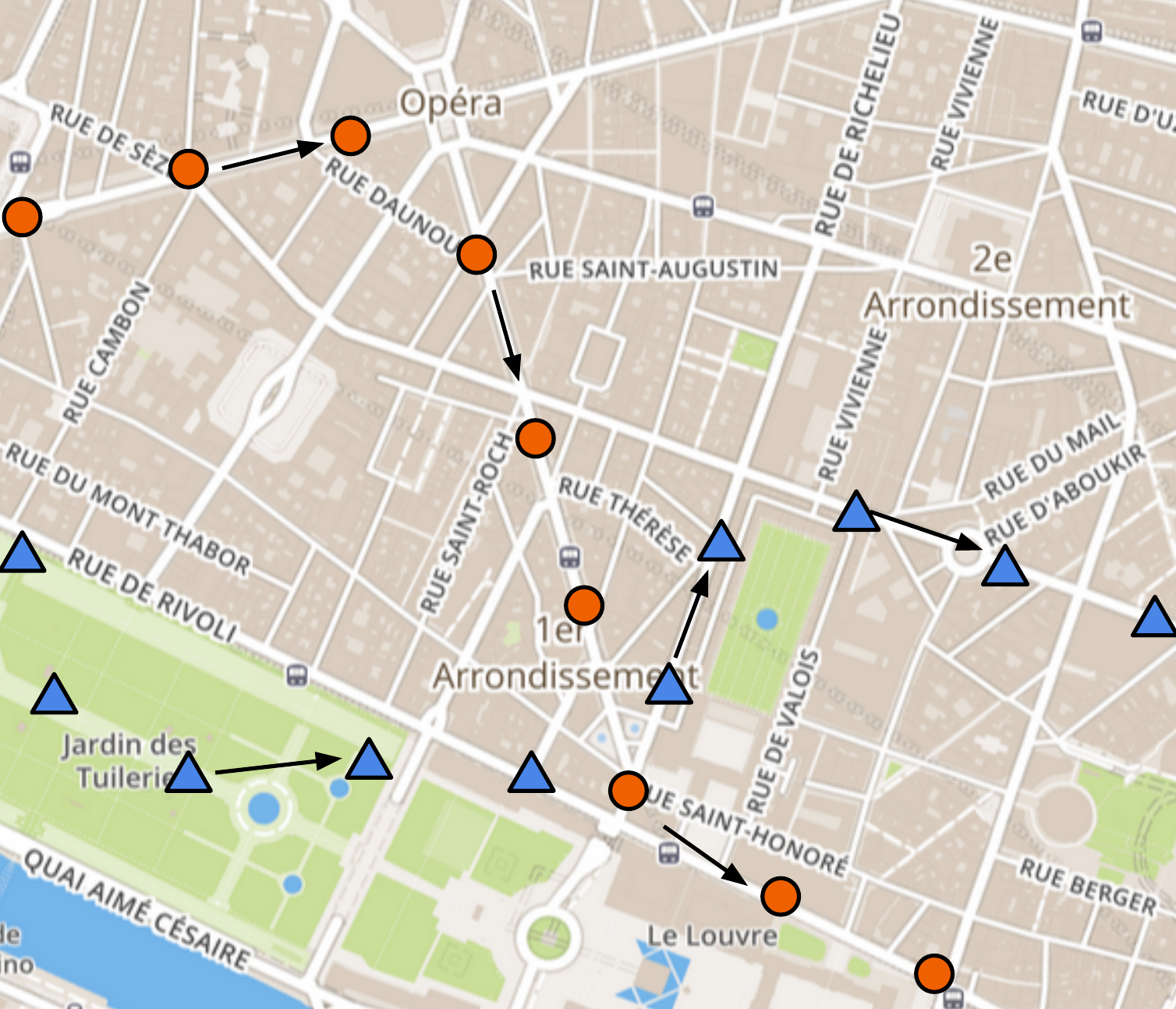}
                \caption{After enforcing a constant speed.}
                \label{fig:smoothing}
        \end{subfigure}
        ~ 
        \begin{subfigure}[b]{0.3\textwidth}
                \includegraphics[width=5.2cm]{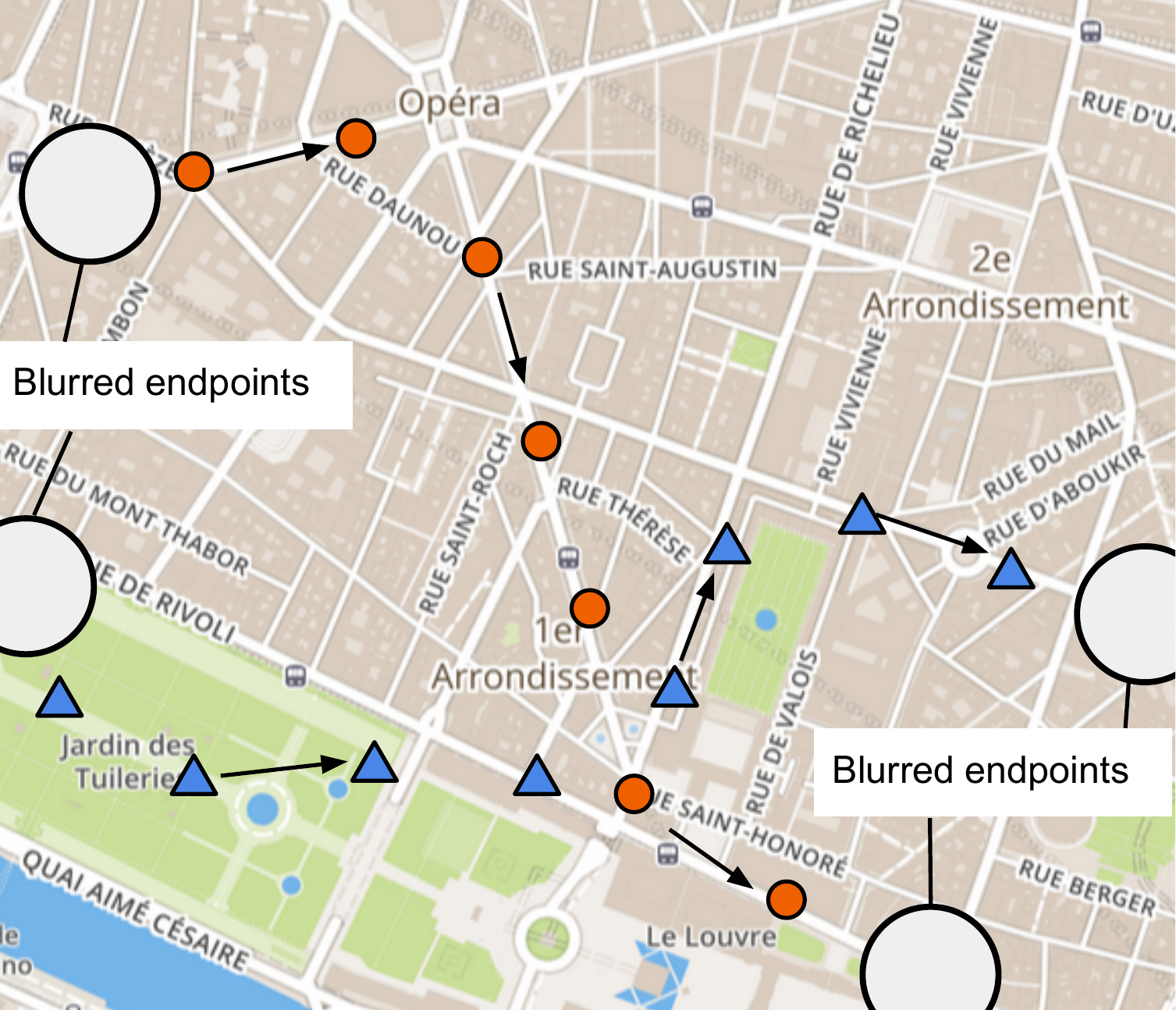}
                \caption{After blurring endpoints.}
                \label{fig:blurred}
        \end{subfigure}
        \caption{Overview of \privapi{}.}\label{fig:overview}
\end{figure*}

\section{A time distortion based mechanism}
\label{sec:detailed}

We begin by proposing definitions of concepts that will be used in the remaining of this paper (Section~\ref{subsec:definitions}), before presenting our mechanism (Section~\ref{subsec:smoothing}).

\subsection{Definitions}
\label{subsec:definitions}

\textbf{Physical and logical user.}
An individual, also called a \textit{physical user}, is a person with an identity (first name, last name, social security number, etc.).
A \textit{logical user} (or \textit{user} for short), is a consistent source of mobility data coming from a single physical user. A physical user can be associated with many logical users (e.g., multiple anonymized traces can be originated from a single physical user).

\textbf{Location.}
A location is a point at the surface of the Earth.
It can be represented in many ways, like by a latitude and a longitude or by a projection in cartesian coordinates.
Locations are elements of $\mathcal{P}$ and we consider that there is a suitable function $d_\mathcal{X} : \mathcal{P}^2 \to \mathbb{R}$ to compute the distance between two locations.

\textbf{Record.}
A record is the smallest indivisible piece of data that can be collected.
It represents the location of a user at a specific time.
More specifically, it is a triplet composed of (1)~a \textit{user identifier}, linking a record to a specific user.
Whereas user identifiers could be anything, we suppose they do not contain anything that could lead to the physical user's identity.
This is why user identifiers are generally sequential or random integers; (2)~a \textit{timestamp}, storing the date and time at which the record was generated; and (3)~a \textit{location}, encoding the spatial representation of the place at which the record was generated.
The set of all possible records being $\mathcal{R}$, we use a dotted notation to ease access to records' members.
For any record $r \in \mathcal{R}$, $r.user$ refers to the user identifier, $r.loc$ refers to the location of the record and $r.time$ refers to the timestamp of the record.

\textbf{Dataset.}
A consistent set of records forms a dataset.
All records inside a dataset usually come from a single collection campaign and feature similar characteristics (e.g., the sampling rate or the geographical area).
The set of all possible datasets being $\mathcal{D}$, for any $D \in \mathcal{D}$, $D \subset \mathcal{R}$.

\textbf{Mobility trace}.
A subset of a dataset that contains all records belonging to a same user is called a mobility trace.
Each user of a dataset is associated with one single mobility trace containing all her records.
The set of all possible mobility traces associated with user $u$ is noted $\mathcal{T}_u$, and hence for any $T \in \mathcal{T}_u$, $T \subset \mathcal{R}$.
There is a total order on $\mathcal{T}_u$ using the chronological order, i.e., more formally, $\forall a, b \in \mathcal{T}_u^2 \, a \leq b \Leftrightarrow a.time \leq b.time$.

\textbf{Location-privacy protection mechanism.}
Datasets can be modified to offer privacy guarantees by using a location-privacy protection mechanism (or simply a \textit{protection mechanism}).
More formally it is defined as a function $\mathcal{D} \longrightarrow \mathcal{D}$ which produces a new dataset offering privacy guarantees from an input dataset.
Protected datasets are also referred to as \textit{published datasets}.
\begin{figure}[h!]
\begin{algorithmic}[1]
    \Function{SmoothSpeed}{$T \in \mathcal{T}_u$, $\epsilon$}
    \State $points \gets \{\}$, $prev \gets \emptyset$
    \For {$curr \in T$}
        \If {$prev = \emptyset$}
            \State $points \gets points \cup \{point\}$
            \State $prev \gets curr.loc$
        \Else
            \State $d \gets d_\mathcal{X}(curr.loc,prev)$
            \While {$d \geq \delta$}
                \State $loc \gets$ \Call{Interpolate}{$prev$, $curr.loc$, $\epsilon$}
                \State $points \gets points \cup \{(loc, curr.time)\}$
                \State $prev \gets loc$
                \State $d \gets d_\mathcal{X}(curr.loc,prev)$
            \EndWhile
        \EndIf
    \EndFor
    \item
    \If {$|points| \leq 2$}
        \State\Return{$\{\}$} \Comment Non protectable trace
    \EndIf
    \item
    \State remove first element of $points$
    \State remove last element of $points$
    \item
    \State $t_{min} \gets \min_{r \in points} r.time$
    \State $t_{max} \gets \max_{r \in points} r.time$
    \State $interval \gets (t_{max} - t_{min})/(|points| - 1)$
    \State $output \gets \{\}$, $time \gets t_{min}$
    \For {$point \in points$}
        \State $output \gets output \cup \{(u, point.loc, time)\}$
        \State $time \gets time + interval$
    \EndFor
    \State \Return{$output$}
    \EndFunction
    \item
    \Function{Promesse}{$D \in \mathcal{D}$, $\epsilon$}
    \State $U \gets \{r.user\ |\ r \in D\}$
    \State $output \gets \{\}$
    \For {$u \in U$}
    	\State $t \gets \{r \in D\ |\ r.user = u\}$
    	\State $output \gets output\ \cup $ \Call{SmoothSpeed}{$t$}
    \EndFor
    \State \Return{$output$}
    \EndFunction
\end{algorithmic}
\caption{\privapi{} implementation.}
\label{alg:speed_smoothing}
\end{figure}

\subsection{Speed smoothing}
\label{subsec:smoothing}

As pointed out in Section~\ref{sec:statement}, our objective in terms of privacy protection is to hide users' POIs.
These correspond to places where users stop and spend some time, before starting again to move to another place.
A trace can be viewed as a list of POIs, that appear on visualized traces as clusters of locations (as shown on Figure~\ref{fig:original}), with transitions in between.
Our counter-measure to hide POIs is thus to enforce a constant speed in the whole trace of a user, i.e., with \textit{speed smoothing}.
If we can guarantee that speed is constant throughout a trace, it becomes difficult for an adversary to spot where a user stopped because there is no point at which she appears to be stationary.
Clues can still be obtained from background knowledge (e.g. the probability is higher to stop in a park that in the middle of a highway) but there will be no certainty for an attacker (e.g. a user can either have just crossed the park or had a picnic there).
Moreover, we want to guarantee that there is a constant duration and distance between two successive records in a trace.
This prevents an attacker from inferring information by studying spatio-temporal intervals at which traces have been sampled.
Figure~\ref{fig:smoothing} shows the result of speed smoothing applied to two mobility traces.
From this figure, we can see that the POIs of the users have been removed and that records on each trace are regularly spaced.
After this step, two records remain unchanged in each trajectory: the first and the last record.
Because they are likely to be POIs (e.g., a home), they need to be protected.
Our solution is to blur endpoints to reduce the precision around them and hence protect users' privacy around these places.
Figure~\ref{fig:blurred} shows the effect of blurring endpoints.

We now present one implementation of the speed smoothing protection mechanism, i.e., \privapi{}, depicted in Figure~\ref{alg:speed_smoothing}.
This algorithm takes as input a mobility trace and an $\epsilon$ parameter.
The first step of this algorithm, is to extract regularly spaced locations, each being at a distance $\epsilon$ from the following one (lines 2-16).
The larger $\epsilon$, the better the privacy guarantee, but the higher the information loss along the trajectory.
To perform this sampling, locations are interpolated along segments joining known locations.
This means that our method is more suited for traces with a high sampling rate (e.g., ten to thirty seconds between consecutive records).
The \textit{Interpolate} function is not given here because its implementation depends on how locations are represented.
After this, we remove the first and last locations from the list (lines 22-23), in order to help hiding endpoints that would otherwise be easily guessed. More precisely, we reduce the precision by $\epsilon$ around there areas.
If after this step there is not enough remaining locations left to recreate a valid trace (i.e., two or less), we simply discard the trace (lines 18-20).
The second step is then to assign to each of the previously sampled locations a timestamp by uniformly distributing the duration of the original trace (lines 25-32).
The implementation of the full \privapi{} protection mechanism is shown in lines 36-44.
The speed smoothing algorithm is independently applied to each mobility trace of a dataset.
The algorithm depicted here is sequential but operations can be parallelized if necessary to anonymize large datasets containing a large number of users.

From the above it appears that a variable that might have an important impact on the results of \privapi{} is $\epsilon$.
This value should be chosen according to the granularity of POIs that the data owner wants to hide.
Intuitively, we expect POIs whose diameter is smaller than $\epsilon$ to be hidden, where the diameter of a POI refers to the diameter of the circular area where all the records related to this POI fall. 
This parametrization question is not unique to our mechanism. Indeed, differentially-private mechanisms~\cite{Dwork06} or $k$-anonymous ones~\cite{Sweeney02} must also be parametrized according to the level of privacy to achieve.


\section{Experimental study}
\label{sec:evaluation}

We start by describing in Section~\ref{subsec:settings} our experimental settings (i.e., datasets, \privapi{} implementation and configuration as well as a description of the other studied mechanisms).
Then, we describe the evaluation of \privapi{} in terms of privacy, utility and performance in Section~\ref{subsec:privacy}, \ref{subsec:utility} and \ref{subsec:performance}, respectively. 

\subsection{Experimental settings}\label{subsec:settings}
\textbf{Datasets.}
We study the protection mechanisms using three real-life datasets: the \textit{Cabspotting} dataset~\cite{Crawdad}, the \textit{Geolife} dataset~\cite{Zheng10} and the \textit{MDC} dataset~\cite{Kiukkonen10,Laurila12}.
Cabspotting has been collected over a month by 536 taxis in the San Francisco Bay Area.
The entire dataset is composed of 11M records.
Geolife has been collected by Microsoft Research Asia over four years and by 182 users.
It is not restricted to people during working hours but instead follows people during their daily life.
This dataset includes 25M records, but is quite inconsistent: some people have been tracked during the whole four years whereas others have only contributed for a few hours.
MDC has been collected between 2009 and 2011 around Lausanne, Switzerland and involves a total of 185 users, followed during their daily life.
The entire dataset is composed of 11M records containing location information.

We pre-processed our datasets to make all traces begin the same day and removed entire days with no data.
Then we only kept the first 20 days of data, to have a dataset with a similar duration for all users.
We further performed another type of pre-processing on the three datasets.
We divided each trace into individual parts, each one being a set of records with no temporal gap between two consecutive records.
Specifically, a trace is divided into two parts when no record is logged during four consecutive hours.
Each part is then considered as an independent trace associated with a new virtual user identifier, no matter to which logical user it really belongs.
This pre-processing helps to preserve privacy, as it breaks the correlation between multiple journeys of a same logical user.
We applied this pre-processing for all studied mechanisms to allow a fair comparison.
Final figures about datasets after the pre-processing are summarized in Table~\ref{tab:datasets}.

\begin{table}
\renewcommand{\arraystretch}{1.3}
\caption{Figures about datasets used in experiments}
\label{tab:datasets}
\centering
\begin{tabular}{l|c|c|c}
\hline
\textbf{Metric} & \textbf{Cabspotting} & \textbf{Geolife} & \textbf{MDC} \\
\hline
Records & 8,986,419 & 3,804,788 & 1,124,454 \\
\hline
Traces & 5,503 & 2,475 & 4,693 \\
\hline
Avg trace duration & 32h 44 min & 3h 4 min & 3h 15 min \\
\hline
Max trace duration & 20 days & 1 day 22h & 20 days \\
\hline
Avg sampling rate & 72s & 7s & 32s \\
\hline
\end{tabular}
\end{table}

\textbf{Implementation.}
Our mechanism is implemented in Java 8.
To enable the reproduction of our experiments, the source code of \privapi{} and datasets we use are available on our website\footnote{\url{http://liris.cnrs.fr/privamov/publications/trustcom15}}. 
We ran our experiments on a single Debian virtual machine having access to 8~Gb of RAM and 8 cores clocked at 1.8~GHz each.

\textbf{Configuration.}
We tested \privapi{} with various values of the $\epsilon$ parameter: 50 (only for Geolife and MDC, because this value is impracticable for Cabspotting), 100, 200 and 500 meters.

\textbf{Other mechanisms studied.}
We compare \privapi{} with two representative state-of-the-art mechanisms.
The first one is Geo-I~\cite{Andres13}, which is the latest approach offering differential privacy guarantees to the users.
We configure Geo-I with various value of $\epsilon$ (despite having the same name, their parameter is not the same as ours and cannot be compared with) that are similar to the ones authors of the paper consider in their publications~\cite{Andres13,Chatzikokolakis14}.
For example, $\epsilon=ln(4)/200$ means that for two points in the radius of 200 meters, their probability to be protected locations of the same original location differ by at most 4.
The lower $\epsilon$, the more noise is added and the stronger the privacy guarantee.
We acknowledge that Geo-I is not exactly designed for the publication of entire datasets, but it is a representative mechanism that adds noise to locations to protect them.
The other mechanism we consider is \wfm{}~\cite{Abul10}, which enforces $k$-anonymity.
We configure \wfm{} to use the LSTD distance, described in their paper, that is shown to perform better with large datasets.
Further, we configure the mechanism with the following parameters: $\delta = 200$ meters, $k = 2$, $Max\_Trash=$ 10~\% of the dataset's size and $max\_radius = 5000$ meters.
This means that at any time, any two traces in the protected dataset are in a cylinder that has a 200 meters diameter.
Other parameters are default ones suggested by the authors of this paper.
We only study one configuration of \wfm{} because $k=2$ is the minimum value (results are worse when $k$ increases~\cite{Abul10}) and $\delta=200m$ puts it in a similar situation than \privapi{}, in addition to being consistent with the value of $\Delta \ell$ we chose (cf. Section~\ref{subsec:privacy}).

\subsection{Privacy evaluation}
\label{subsec:privacy}
We evaluate the privacy effectively guaranteed to users by running a POIs extraction attack on datasets protected by \privapi{}, Geo-I and \wfm{}.
Using the result of an attack to quantify privacy is an approach that has already been proposed in~\cite{Shokri11}.
POIs can be extracted using algorithms such as the one proposed in~\cite{Hariharan04}.
We configure this algorithm to extract POIs with a maximum diameter $\Delta \ell = 200$ meters and a minimum stay time $\Delta t = 15$ minutes.
To measure the privacy leakage, we extract POIs from the dataset before and after having applied a protection mechanism and whether the latter match with the former.
Specifically, we note $P \in \mathcal{P}^n$ a set of POIs extracted from an original trace and $P' \in \mathcal{P}^{n'}$ a set of POIs extracted from a protected trace (POIs can be viewed as simple locations).

\begin{definition} The matching operation associates to each location of a given set, the closest location in another set iff its distance w.r.t. $d_\mathcal{X}$ is lower than a threshold $\ell$:

\[match_\ell(P,P') = \bigcup\limits_{p' \in P'} arg \min\limits_{p \in P} \{ d_\mathcal{X}(p, p') | d_\mathcal{X}(p, p') \leq \ell \}.\]
\end{definition}

\begin{definition} The recall is the ratio between the number of matching POIs retrieved from a protected dataset and the real number of POIs:

\[recall_\ell(P, P') = \frac{|P \, \cap \, match_\ell(P,P')|}{|P|}.\]
\end{definition}

\begin{definition} The precision is the ratio between the number of matching POIs retrieved from a protected dataset and the total number of POIs retrieved:

\[precision_\ell(P, P') = \frac{|P \, \cap \, match_\ell(P,P')|}{|P'|}.\]
\end{definition}

\begin{definition} The F-score is the harmonic mean of precision and recall:

\[fscore_\ell(P, P') = \frac{2 \times precision_\ell(P,P') \times recall_\ell(P,P')}{precision_\ell(P,P') + recall_\ell(P,P')}.\]
\end{definition}


\begin{table}[htb]
\renewcommand{\arraystretch}{1.3}
\caption{F-scores of POIs retrieval.}
\label{tab:fscore}
\centering
\begin{tabular}{l|c|c|c}
\hline
\textbf{Protection mechanism} & \textbf{Cabspotting} & \textbf{Geolife} & \textbf{MDC} \\
\hline
\privapi{}, $\epsilon=50m$ & - & 17.22~\% & 13.91~\% \\
\hline
\privapi{}, $\epsilon=100m$ & 0~\% & 11.06~\% & 7.94~\% \\
\hline
\privapi{}, $\epsilon=200m$ & 0~\% & 2.27~\% & 1.01~\% \\
\hline
\privapi{}, $\epsilon=500m$ & 0~\% & 0~\% & 0~\% \\
\hline\hline
Geo-I, $\epsilon=ln(10)/100$ & 30.72~\% & 42.38~\% & 55.95~\% \\
\hline
Geo-I, $\epsilon=ln(6)/200$ & 6.33~\% & 9.41~\% & 16.31~\% \\
\hline
Geo-I, $\epsilon=ln(4)/200$ & 3.53~\% & 4.50~\% & 9~\% \\
\hline
Geo-I, $\epsilon=ln(2)/200$ & 1.49~\% & 0.85~\% & 0.54~\% \\
\hline\hline
\wfm{}, $k=2$, $\delta=200m$ & 0~\% & 0~\% & 0~\% \\
\hline
\end{tabular}
\end{table}

For conciseness, we only report on the F-score metric with $\ell = \Delta \ell / 2 = 100m$, which takes into account both precision and recall.
We report on the average F-score across all traces in each dataset in Table~\ref{tab:fscore}.
From this table, we can see that in the Cabspotting dataset, POIs are always hidden with \privapi{}, no matter the value of $\epsilon$.
With Geolife and MDC, we closely reach our goal with $\epsilon=200m$.
This value of  $\epsilon$ was indeed expected to be its optimal parametrization, given that with this value we have $\epsilon = \Delta \ell$.
Still, the F-Score is close but not equal to zero, most likely due to some edge cases.
It is worth noting that behind this low F-Score, we have in reality both a low precision and a low recall.
This means that very few POIs are retrieved, and that they are lost inside many false positives.
Furthermore, we observe that \wfm{} hides all POIs in the three datasets, but this is a consequence of the great quantity of noise that has to be added to enforce $k$-anonymity.
Here the privacy comes at the cost of a very degraded utility, as it will be shown in Section~\ref{subsec:utility}.
Finally, retrieving POIs from the Geo-I protected datasets is very dependent on the quantity of noise that has been added.
With lower values of $\epsilon$, and hence more privacy, retrieving POIs becomes very challenging.
With $\epsilon=ln(2)/200$, almost no POI is found, as with \privapi{}.
Weaker values of $\epsilon$ allow many POIs to be retrieved, up to an F-Score of 55~\%.
We stayed with a simple POIs extraction routine, but risk is even higher if we enlarge the clustering radius $\Delta \ell$ to adapt to Geo-I, as it has been shown in~\cite{Mapomme14}.

\subsection{Utility evaluation}
\label{subsec:utility}

We evaluate the utility of the protected datasets with two metrics: the spatial error and range queries as described below.

\textbf{Spatial error.}
We compute the difference between two traces by measuring the distance between each record in the protected trace and its projection onto the original trace.
We consider original traces as polylines whose vertices are records.
This allows us to fill the gaps between records and project points onto these polylines.
The distance between a record in the protected trace and its projection onto the original trace represents the spatial error, i.e., the spatial inaccuracy of a protection mechanism.
Please note that we do not take into account the error due to the interpolation between two consecutive locations, which is what our mechanism actually does.
We consider the original trace to be precise enough to allow interpolation without losing accuracy.

\begin{table}[htb]
\renewcommand{\arraystretch}{1.3}
\caption{Spatial errors.}
\label{tab:locerror}
\centering
\begin{tabular}{l|c|c|c}
\hline
\textbf{Protection mechanism} & \textbf{Cabspotting} & \textbf{Geolife} & \textbf{MDC} \\
\hline
\privapi{}, $\forall \epsilon$ & 0 m & 0 m & 0 m \\
\hline\hline
Geo-I, $\epsilon=ln(10)/100$ & 24 m & 45 m & 52 m \\
\hline
Geo-I, $\epsilon=ln(6)/200$ & 50 m & 120 m & 140 m \\
\hline
Geo-I, $\epsilon=ln(4)/200$ & 62 m & 156 m & 183 m \\
\hline
Geo-I, $\epsilon=ln(2)/200$ & 113 m & 325 m & 378 m \\
\hline\hline
\wfm{}, $k=2$, $\delta=200m$ & 13,046 m & 69,676 m & 19,222 m \\
\hline
\end{tabular}
\end{table}

We report about the average spatial error for all records in each dataset in Table~\ref{tab:locerror}.
From this table, we observe that the spatial error of \privapi{} is equal to zero for the three datasets.
Indeed, by construction, the only inaccuracy introduced by \privapi{} is due to the interpolation between sampled records, which shows to be negligible in this experiment.
Geo-I instead adds noise to locations, depending on its $\epsilon$ parameter, which results in an average error ranging from 24 to 378 meters on the three datasets.
This has to be compared with the average error due to GPS measurements which is about 15-20 meters.
This means that at its weakest level of privacy, Geo-I is just a little bit less precise than the error that could come from the  normal usage of a GPS.
However, when the level of privacy increases, the error can go as high as 378 meters, which is enough to highly disturb data mining tasks, especially in a dense urban environment.
Finally, among the three tested mechanisms, \wfm{} is the one with the worst spatial error, which is at least equal to 13,046 meters in our experiments. This is due to the large amount of noise \wfm{} introduces to enforce k-anonymity.
These results highlight the benefit of a time distortion protection mechanism for use cases where a high spatial accuracy is needed, which cannot be achieved with the other mechanisms building on spatial distortion.

\textbf{Range queries.}
A classical operation performed on published datasets is \textit{range queries}.
A range query counts how many unique (i.e., different) users cross an area during a time window.
Despite being simple, this operation can be employed by many useful applications (e.g., traffic prediction, finding popular places).
To measure the utility related to range queries, we use the \textit{range query distortion} metric defined in~\cite{Abul10}.

\begin{definition}
The distortion associated with a range query $Q$ is the relative error between its result over the original dataset $D \in \mathcal{D}$ and its result over a protected dataset $D' \in \mathcal{D}$:

\[distortion_Q(D,D') = \frac{|Q(D) - Q(D')|}{Q(D)}.\]
\end{definition}

In our context, a range query is defined using two parameters: a time window and a geographical area.
Because the set of all range queries is infinite, we need to generate many range queries with different time windows and areas centered around different times and locations.
Additionally, many range queries are irrelevant for our experiments, because they concern areas or times for which we have no data.
This is why we use a smart range query generator choosing randomly one record in the evaluated original dataset and using it to generate a query centered around this record with a random duration and a random area. 
This ensures that $Q(D) \neq 0$ (and that our distortion is always defined).
Similarly to~\cite{Abul10}, we choose time windows ranging from 2 hours to 8 hours and square areas whose half-diagonals range from 500 to 5,000 meters.

\begin{table}[htb]
\renewcommand{\arraystretch}{1.3}
\caption{Range query distortions.}
\label{tab:distortion}
\centering
\begin{tabular}{l|c|c|c}
\hline
\textbf{Protection mechanism} & \textbf{Cabspotting} & \textbf{Geolife} & \textbf{MDC} \\
\hline
\privapi{}, $\epsilon=50m$ & - & 15.14~\% & 25.4~\% \\
\hline
\privapi{}, $\epsilon=100m$ & 6.77~\% & 14.83~\% & 25.23~\% \\
\hline
\privapi{}, $\epsilon=200m$ & 6~\% & 15.1~\% & 27~\% \\
\hline
\privapi{}, $\epsilon=500m$ & 6.9~\% & 18.97~\% & 31.23~\% \\
\hline\hline
Geo-I, $\epsilon=ln(10)/100$ & 0.7~\% & 7.58~\% & 3.08~\% \\
\hline
Geo-I, $\epsilon=ln(6)/200$ & 2.46~\% & 20.43~\% & 10.36~\% \\
\hline
Geo-I, $\epsilon=ln(4)/200$ & 3.39~\% & 27.24~\% & 13.47~\% \\
\hline
Geo-I, $\epsilon=ln(2)/200$ & 7.21~\% & 60.46~\% & 29.98~\% \\
\hline\hline
\wfm{}, $k=2$, $\delta=200m$ & 102~\% & 102~\% & 94~\% \\
\hline
\end{tabular}
\end{table}

We report about the average query distortion in Table~\ref{tab:distortion}, which is the average distortion over 1,000 randomly generated queries.
Results show that \privapi{} has a query distortion ranging from 6~\% to 27~\% for $\epsilon=200m$.
This means that results of range queries have, on average, a relative error of at most 27~\%.
Further, results show that we perform at least 71~\% better than \wfm{} with all our datasets.
Once again, the distortion with Geo-I is dependent on the value of $\epsilon$. The weakest value features almost no distortion (but also does not protect POIs in a satisfactory way, cf. previous section).
Intermediary values of $\epsilon$ correspond to similar distortions than with the optimal \privapi{}.
Note that the range query distortion metric is more sensitive with Geolife and MDC because counts are way smaller than with Cabspotting, and therefore the effect of missing one user is more important on the relative error.
This explains the larger difference of distortion between Geo-I and \privapi{} on the Geolife and MDC datasets.

\subsection{Performance evaluation}\label{subsec:performance}
We also consider two non-functional metrics that can also impact the usability of the protected datasets, i.e., the execution time and the compression degree as described below.

\textbf{Execution time.}
The execution time is the time taken by the anonymization process to generate a new protected dataset.
We assume the original dataset is already stored in the file system (e.g., a database or a distributed file system) and only measure the time to generate and write the protected dataset in a text file.
We report about execution times in Table~\ref{tab:execution}.
Because execution times remain almost constant for the various configurations of Geo-I and \privapi{}, we do not report about subtle variations depending on parameters.
Geo-I and \privapi{} are the fastest mechanisms because their algorithms are quite simple.
Geo-I independently adds noise to each record and \privapi{} independently protects each trace, which enables them to be efficient.
\wfm{} is the slowest mechanism because of the complexity coming from the clustering of similar traces that is the heart of this mechanism. 

\begin{table}[hbt]
\renewcommand{\arraystretch}{1.3}
\caption{Execution times.}
\label{tab:execution}
\centering
\begin{tabular}{l|c|c|c}
\hline
\textbf{Protection mechanism} & \textbf{Cabspotting} & \textbf{Geolife} & \textbf{MDC} \\
\hline
\privapi{} & 210 s & 25 s & 15 s \\
\hline
Geo-I & 147 s & 64 s & 21 s \\
\hline
\wfm{} & 605 min & 70 min & 37 min \\
\hline
\end{tabular}
\end{table}

\textbf{Compression.}
In a general context, the anonymized dataset can have a different size than the original dataset.
It can be smaller, if records have been deleted, or larger, if fake records have been added.
Producing datasets that are orders of magnitude larger than the original ones greatly decreases their usability, because the time needed to load and query them increases accordingly, while much smaller ones can introduce information losses (that could be quantified with the previous utility metrics).

\begin{definition}
We define the compression degree as the ratio between the size of a protected dataset $D' \in \mathcal{D}$ and the size of the non-empty original dataset $D \in \mathcal{D}$:

\[compression(D, D') = \frac{|D'|}{|D|}.\]
\end{definition}

\begin{table}[htb]
\renewcommand{\arraystretch}{1.3}
\caption{Compression degrees.}
\label{tab:compression}
\centering
\begin{tabular}{l|c|c|c}
\hline
\textbf{Protection mechanism} & \textbf{Cabspotting} & \textbf{Geolife} & \textbf{MDC} \\
\hline
\privapi{}, $\epsilon=50m$ & - & 51~\% & 206~\% \\
\hline
\privapi{}, $\epsilon=100m$ & 369~\% & 25~\% & 101~\% \\
\hline
\privapi{}, $\epsilon=200m$ & 177~\% & 12~\% & 48~\% \\
\hline
\privapi{}, $\epsilon=500m$ & 64~\% & 4~\% & 18~\% \\
\hline\hline
Geo-I, $\forall \epsilon$ & \multicolumn{3}{c}{100~\%} \\
\hline\hline
\wfm{}, $k=2$, $\delta=200m$ & 106~\% & 76~\% & 99~\% \\
\hline
\end{tabular}
\end{table}

Experimental results for the compression are shown in Table~\ref{tab:compression}.
Results show that because Cabspotting has a coarser sampling rate than Geolife, for small values of $\epsilon$ the protected dataset is much larger than the original one with \privapi{}.
Conversely the Geolife dataset protected with \privapi{} is much smaller than the original one, the original dataset having been collected with a very high sampling rate (average sampling rate is 7 seconds, cf. Table~\ref{tab:datasets}).
\wfm{} and Geo-I both have almost no effect on the size of the produced dataset.
This metric is interesting because it shows that for \privapi{}, some values of $\epsilon$ are impracticable, resulting in too huge datasets (this is why we do not experiment with Cabspotting at $\epsilon=50m$).
But it also shows that it is possible to reduce the size of a dataset with a high sampling rate without losing "too much" information (cf. range query distortion).

\subsection{Summary of the study}
From the results presented in this section, we conclude that time distortion seem to be a promising alternative to spatial distortion for the privacy-preserving publication of mobility datasets.
Indeed, on the three datasets we studied, our proposed \privapi{} mechanism parametrized with $\epsilon=200m$ hides almost all users' POIs, while keeping the spatial accuracy very high.
Temporal distortion has though an impact on metrics for which time is important, e.g., range queries.
Nevertheless, \privapi{} still offers a distortion varying from 6~\% to 27~\% according to the sparsity of dataset, which is comparable to Geo-I and way better than \wfm{}.
Furthermore, \privapi{} is simple to parametrize because there is only one parameter $\epsilon$ to set, whose meaning is clear: it represents the granularity of POIs to protect. 
Obviously, among the many use cases that data analysts may want to implement, there shall be some that require a high temporal accuracy, which \privapi{} cannot provide.
In this case, Geo-I may still be a better candidate. 
Our goal in this paper was to introduce another way to anonymize mobility data, that takes the opposite direction of actual state-of-the-art protection mechanisms and to practically study its effectiveness.

\section{Conclusion}
\label{sec:conclusion}

In this paper we presented \privapi{}, a new protection mechanism to anonymize mobility datasets.
Its novelty resides in the fact that it distorts time instead of distorting location, which allows it to have a better utility than representative state of the art mechanisms.
Privacy evaluation shows that, when configured appropriately, \privapi{} resists POI discovery attacks similarly to Geo-I, which enforces differential privacy.
\wfm{}, which enforces $k$-anonymity, still performs better, but at the cost of a very decreased utility.
Finally, \privapi{} is fast, as it can anonymize a dataset of 9M records in less than four minutes.
From our study, we conclude that time distortion anonymization is a promising alternative to existing spatial ones, particularly for use cases where a high spatial accuracy is required.
In our future work, we plan to improve the implementation of our algorithm by producing more realistic traces without impacting the privacy guarantees.

\section*{Acknowledgment}
This work was supported by the LABEX IMU (ANR-10-LABX-0088) of Université de Lyon, within the program "Investissements d'Avenir" (ANR-11-IDEX-0007) operated by the French National Research Agency (ANR).
Portions of the research in this paper used the MDC Database made available by Idiap Research Institute, Switzerland and owned by Nokia.

\bibliographystyle{IEEEtran}
\bibliography{IEEEabrv,bibli}

\end{document}